# Reconstructing the Solar Wind From Its Early History to Current Epoch

*Vladimir S. Airapetian[1] and Arcadi V. Usmanov[2]*
[1]*NASA/GSFC,* [2]*University of Delaware & NASA/GSFC*

**Abstract.**

Stellar winds from active solar-type stars can play a crucial role in removal of stellar angular momentum and erosion of planetary atmospheres. However, major wind properties except for mass loss rates cannot be directly derived from observations. We employed a three-dimensional magnetohydrodynamic Alfvén wave driven solar wind model, ALF3D, to reconstruct the solar wind parameters including the mass loss rate, terminal velocity and wind temperature at 0.7, 2 and 4.65 Gyr. Our model treats the wind thermal electrons, protons and pickup protons as separate fluids and incorporates turbulence transport, eddy viscosity, turbulent resistivity, and turbulent heating to properly describe proton and electron temperatures of the solar wind. To study the evolution of the solar wind, we specified three input model parameters, the plasma density, Alfvén wave amplitude and the strength of the dipole magnetic field at the wind base for each of three solar wind evolution models that are consistent with observational constrains. Our model results show that the velocity of the paleo solar wind was twice as fast, ~ 50 times denser and 2 times hotter at 1 AU in the Sun's early history at 0.7 Gyr. The theoretical calculations of mass loss rate appear to be in agreement with the empirically derived values for stars of various ages. These results can provide realistic constraints for wind dynamic pressures on magnetospheres of (exo)planets around the young Sun and other active stars, which is crucial in realistic assessment of the Joule heating of their ionospheres and corresponding effects of atmospheric erosion.

## 1. Introduction.

The Sun, a typical G2 V star, exhibits magnetic activity in the form of magnetically driven eruptive processes including flares, solar wind, coronal mass ejections (CME) and associated solar energetic particle events. The inputs of these forms of solar activity drive space weather by energizing the environment around the Sun and Earth. Earth-directed energetic CMEs cause magnetospheric perturbations and may induce large geomagnetic currents (Schrijver et al. 2014). The observations of young stars resembling our Sun in its infancy suggest that the magnetic activity and associated paleo space weather events were much more frequent and violent and the solar wind was much denser in the Sun's past (Wood et al. 2005; Airapetian et al. 2015a). Winds from young stars can serve as an efficient mechanism in removing angular momentum and thus causing their host stars to spin down with time (Skumanich 1972). The solar and stellar winds may also have directly contributed to the atmospheric erosion, and in this way impacted climate and planetary habitability. Recent measurements of atmospheric loss in planetary atmospheres suggest that increase of the solar wind dynamic pressure by a factor of two enhances the mass loss of $O^+$ ions up to 10 times for Earth and 60 times for Mars (Wei et al. 2012). Thus, the dynamic pressure exerted by denser and faster paleo solar wind should be considered as one of the major factors affecting the erosion of the atmospheres of early Venus, Mars, Earth, Earth twins and hot Jupiters (Lammer et al. 2003; Griessmeier et al. 2004; Kulikov et al. 2007; Lundin et al. 2007; Khodachenko et al. 2012; Kislyakova et al. 2015). It also affects the size, density and magnetic field at the heliospheric boundary (heliopause) modulating the flux of Galactic cosmic rays, which may cause changes in cloud cover and ozone destruction, and therefore, could influence habitability conditions on early Earth and Mars (Pavlov et al. 2005; Usoskin & Kovaltsov 2008; Potgieter et al. 2013).

In order to characterize the dynamic pressure, the mass loss rate and velocity of the paleo solar wind need to be reconstructed. This can be accomplished by observing the signatures of winds from young solar analogs (Ribas et al. 2005). However, unlike the solar wind, the properties of stellar winds are poorly known, because very few observational constrains are available. Currently, the mass loss rates from the solar-like stars can be derived from the detection of stellar "astrospheres" forming due to interaction of stellar winds with the interstellar medium (Wood et al. 2005). These estimates imply that mass-loss rate, $\dot{M}$, declines with age of solar-type stars, as $t^{-2.33\pm0.55}$ due to angular momentum loss caused by magnetized winds. This method is limited to stars with ages ≥ 0.7 Gyr due to the saturation of the X-ray emission for young stars. The application of this technique for young stars implies that the mass-loss rate from the paleo solar wind at 0.7 Gyr, the time when life started on Earth, was ~100 times greater than the one observed today. However, this method assumes a spherically symmetric wind with the speed of 400 km/s wind speed for all astrospheric models, which gives n uncertainty in the determination of the mass loss rate.

Another way to characterize the paleo solar wind is to apply the wind models of the current Sun with the input conditions relevant for the young and active solar-type stars. Remote sensing and in-situ measurements suggest that near solar minimum the solar wind has a bimodal structure with (1) a high-speed (up to 800 km/s at 1 AU), low density (1-2 cm$^{-3}$) component emanating from the polar regions associated with unipolar coronal holes and (2) a wind about 5 times denser with half the fast wind's speed (~400 km/s) that forms above the low latitude regions associated with large-scale bipolar structures known as coronal streamers (McComas et al. 2003). Parker's thermally driven solar wind model (Parker, 1958) with T=1-2 MK can well reproduce the slow component of the solar wind, but fails to explain the fast wind component. Alfvén waves are believed to provide the energy and momentum to heat the chromosphere and drive the fast solar wind (Ofman and Davila 1998; Usmanov et al. 2000; McIntosh et al. 2011; Cranmer & Saar 2011; van der Holst et al. 2010; Suzuki et al. 2013; Matsumoto & Suzuki 2014; Airapetian and Cuntz 2015).

One class of solar/stellar wind models is based on scaling between wind properties and stellar parameters including age and rotation rate using simple analytical and one-dimensional models (Holzwarth and Jardine 2007; Cranmer and Saar 2011; Matt et al. 2012; Johnstone et al. 2015). Another class of models utilizes the multi-dimensional MHD treatment (Vidotto et al. 2009; Sterenborg et al. 2011; Cohen 2011; Cohen and Drake 2014; Matsumoto & Suzuki 2014). Sterenborg et al. (2011) applied a 3D MHD model to study the mass loss rate and velocity of the thermally driven paleo solar wind by varying density and magnetic field of the coronal wind base. They simulated the slow component of the paleo wind and concluded that in order to obtain the mass loss rate of 100 times the current Sun's rate, a high base density of 10$^{10}$ cm$^{-3}$ is needed.

In this paper we report our first results from simulations of fast and slow solar wind components throughout the Sun's evolution using a three-fluid, three-dimensional magnetohydrodynamic code, ALF3D (Usmanov et al. 2014). Unlike previous models, our global model describes the solar wind in the region that extends from the coronal base to 100 AU and includes effects of thermal conduction and Alfvén wave pressure as well as plasma heating due to turbulent dissipation. The model treats the wind thermal protons, electrons, and pickup protons as separate fluids and includes turbulence transport, eddy viscosity, turbulent resistivity, and turbulent heating (Usmanov et al. 2014). In Section 2, we describe model parameters applicable for the young Sun's coronal conditions. In Section 3, we present and discuss the model results, and draw our conclusions in Section 4.

## 2. Setup Model Parameters.

We employ a three-fluid 3D MHD simulation code, ALF3D, and the method described in Usmanov et al. (2014) to compute steady-state wind solutions for three wind scenarios: the young (0.7 Gyr), intermediate age (2 Gyr) and the current Sun from the coronal wind base to 3 AU. The solar wind is described as a three-fluid plasma containing thermal solar wind electrons and protons and interstellar pickup protons. We treat the three plasma components by separate energy equations and assume that the bulk wind velocity is the same for these species (Usmanov et al. 2014). The code solves fully three-dimensional time-dependent compressible MHD equations under the assumption of dipole magnetic filed in the corotating frame of reference. The magnetic dipole is assumed to be aligned with the rotation axis. For each of the three evolutionary scenarios, we assume that the solar wind is driven by a combination of the thermal pressure gradient (Parker 1958) and Alfvén waves generated at the coronal wind base. The initial and boundary conditions for the current Sun simulations are similar to those described in Usmanov et al. (2014). The computational domain in our wind model is sub-divided into three separate regions. The first, inner wind region, extends from 1 to 20 $R_\odot$ ($R_\odot$ is the solar radius). In this region, the solar corona is described in the single-fluid polytropic approximation, with the polytropic index γ=1.08 or as nearly isothermal plasma. The outer boundary of the inner region is set at 20 $R_\odot$ to assure that the solar wind flow at the boundary is supersonic and super-Alfvénic. We obtain a steady-state solution for the transonic and trans-Alfvénic single-fluid coronal wind by using the time-relaxation method described by Usmanov et al. (2000). The inner boundary conditions specified at the wind base, $r = 1R_\odot$, are updated in the course of the relaxation process to be consistent with the flow characteristics near the coronal boundary (Usmanov et al. 2000). They evolve to a steady state simultaneously with the solution in the inner region. The outflow boundary conditions at 20 $R_\odot$ are approximated by a first-order (linear) extrapolation. In the second region extending between 20 $R_\odot$ and 0.3 AU, the solar wind flow is supersonic and super-Alfvénic. In this region, the adiabatic index γ = 5/3 is used in combination with Hollweg's electron heat flux (Hollweg, 1974, 1976). Because plasma in the second region (at heights > 20 $R_\odot$) is collisionless, we apply a two-fluid approximation including thermal electrons and protons to describe the solar wind. A steady-state solution in this region is constructed using a marching-along-radius numerical method (Pizzo 1978, 1982; Usmanov 1993).

The third, outer wind region starts at 0.3 AU and covers the heliosphere up to 100 AU, where plasma is treated as three-fluid plasma composed of electrons, thermal protons and pickup protons. This is critical in correctly calculating the wind velocity due to interaction with the incoming interstellar hydrogen. In this region, in order to describe the Alfvén wave driven solar wind turbulence and associated plasma heating rate, we decompose plasma quantities including plasma pressure, velocity and magnetic field into the Reynolds averaged and fluctuating components (see Usmanov et al. 2014) and use the turbulence transport equations with an eddy-viscosity approximation for the Reynolds stress tensor and turbulent electric field. We then obtain steady-state solutions using time relaxation that incorporates boundary conditions from the intermediate (the second) region.

The current approach overcomes an important limitation of the solar and stellar wind models of Usmanov et al. (2012) and Cohen and Drake (2014) that describe the solar wind at large distances from its base as a single fluid. The inclusion of pickup protons treated as a separate fluid affects the kinematics of the solar wind at distances greater than 5-10 AU due to the momentum transfer from the solar wind protons to pickup protons. This also accounts for solar wind plasma heating due to dissipation of turbulent wave energy, the compression of the predominantly azimuthal magnetic field in the outer heliosphere as a result of the deceleration, and the weakening of the corotating interaction regions caused by the deceleration and the pressure of pickup protons. The inclusion of pickup protons is especially important in order to describe the wind properties and its thermodynamics at greater distances from the Sun and to properly characterize the dynamic pressure of the solar wind at various epochs in the Sun's history.

We applied our solar wind model in the region from the coronal wind base to 3 AU for three wind evolutionary scenarios, M1, M2 and M3, reproducing the conditions in the early Sun at 0.7 Gyr with the rotation period of 5 days (at equator), the intermediate-age Sun at 2 Gyr with the rotation period of 10 days, and the wind model for the current Sun with the rotation period of 25 days (see Table 1). The M1 scenario is described by the observed parameters of the young solar analog, $k^1$ Cet, as a representative case for the M1 scenario. The age of this star, 0.7 Gyr was derived using the scaling law $\tau \sim L_x^{-0.67}$ (Güdel et al. 1997), where $L_x$ is its X-ray luminosity (Ness et al. 2004). To characterize the star's coronal density at the wind base, we use the emission measure of the coronal emission OVII triplet that is formed in the low corona. The density is at least one order of magnitude greater than the full disk Sun's (the Sun as a star) coronal density (Laming et al. 1995; Ness et al. 2004). The term coronal density refers to the density associated with coronal magnetic loops that are much denser, and therefore, brighter than the magnetically open regions associated with the solar wind. While there are no direct density measurements from stellar coronal holes, we assume that the wind base density in the star's polar region is about 10-12 times larger than the solar wind base density, $N_b = 4 \times 10^7$ cm$^{-3}$ used for M1, the current solar wind scenario (Usmanov et al. 2014).

The M2 wind scenario is represented by the properties of the intermediate age (~2 Gyr) solar-like K2V star, ε Eri, with the rotation period of 15 days. Its coronal density derived from the OVII triplet is $10^{10}$ cm$^{-3}$, or ~4 times the full disk Sun's coronal density (Doyle 1980; Ness et al. 2002). We use this scaling factor to set the density of the wind base for the M2 scenario (see Table 1). The M3 solar wind scenario is identical to the "solar minimum" model presented in Usmanov et al. (2014).

| Model | Rotation period (in days) | $N_b$ (in $10^7$ cm$^{-3}$) | $B_0$ (G) | $\delta V$ (in km/s) | $T_b$ (K) | $\dot{M}$ (in $M_\odot$/yr) |
|---|---|---|---|---|---|---|
| M1 (Sun at 0.7Gyr, $k^1$ Cet) | 5 | 50 | 54.9 | 100 | 1.6/1.8/2.0 | 26/37/50 |
| M2 (Sun at 2 Gyr, ε Eri) | 10 | 17 | 27.6 | 50 | 1.6/1.8/2.0 | 3.2/6/10 |
| M3 (Sun at 4.65 Gyr) | 25 | 4 | 16 | 35 | 1.8 | 1 |

Table 1. Initial and final wind properties for the M1, M2 and M3 solar wind scenarios. To study the effect of the uncertainty of the input coronal base temperature, $T_b$, we present three values of the mass loss rates, $\dot{M}$, for three values of $T_b$.

While the coronal emission measure distribution over temperatures can be retrieved directly from optically thin lines forming at the representative range of coronal temperatures, the stellar coronal temperature at the wind base cannot be determined directly unless we have a physically justified model of the coronal base. Unfortunately, currently no physical model of the chromospheric and transition region heating can be directly applicable to predict the heating rates, density, temperature distribution and the Alfvén wave flux at the top of the transition region associated with the wind base (Airapetian and Cuntz 2015). The existing models can describe this atmospheric properties only in a parameterized way (Suzuki et al. 2013; Matsumoto & Suzuki 2014), because a realistic treatment of the solar chromosphere should include an assessment of Joule and viscous heating introduced by wave dissipation, as well as conductive cooling and a detailed treatment of radiative transfer that is important in calculating the radiative cooling rate. We have started the first step in this direction by studying the heating rates associated with wave dissipation in a partially ionized stellar chromosphere caused by upward propagating and reflected Alfvén waves launched at the photosphere (Airapetian et al. 2015b).

In this paper we assume that the wind originates at the low coronal base and we calculate each scenario for three wind base temperatures, 1.6, 1.8 and 2.0 MK. This is consistent with arguments by Matt et al. (2012) that the wind base temperature does not depend on age (or stellar rotation rate), but is determined by only stellar mass and radius.

The solar dynamo operates by alternating between solar poloidal (or dipole-like) and toroidal field (Parker 1979). During solar minimum the Sun's magnetic field can be adequately described as a dipole that gives way to higher harmonics during solar maximum. In the current study, we restrict ourselves with the assumption of a simple dipole magnetic field with no tilt with respect to the solar rotation axis in all three models, M1-M3. The magnetic dipole is specified by the strength of the magnetic field on the solar pole, $B_0$. The value for $B_0$ in the M3 model was selected to match the in-situ measurements of the interplanetary wind magnetic field at 1 AU (Zhao and Hoeksema 1995). The dipolar field strength for earlier epochs (M1 and M2 models) was selected to match the statistical correlation between the average surface field and the stellar age derived by the "Bcool Collaboration" team (Vidotto et al. 2014) as <$B_V$> ~ $t^{-0.655 \pm 0.045}$.

Our model of Alfvén wave driven wind uses the amplitude of Alfvén waves at the coronal wind base. This value can be derived directly from spectroscopic observations assuming that the observed non-thermal broadening of optically thin emission lines is caused by turbulent motions driven by Alfvén waves along the line of sight. The initial wave amplitude at the wind base, δV, is chosen to be consistent with the observationally derived values for the current Sun and young solar-like stars (Hassler et al. 1990; Wood et al. 1997). The typical value for the Alfvén wave amplitudes of 35 km/s specified for the solar coronal base is derived from the non-thermal broadenings of the Mg X, Fe XII, OV and NV emission lines observed above the limb. The non-thermal broadening is a universal signature of transition region and coronal emission lines observed in active stars, and its magnitude increases with the activity level and stellar age. Intermediated-aged stars show line broadenings ~50 km/s, while this amplitude reaches over 100 km/s for young stars (Wood et al. 1997; Brandt et al. 2001; Ayres 2015). The knowledge of the density, magnetic field and the wave amplitude, δV, sets the total energy flux in Alfvén waves, ρ δV² $V_A$, propagating upward into the solar corona and accelerating solar winds.

## 3. Simulation Results.

Because of the uncertainties in our knowledge of the wind base density, temperature, magnetic field and Alfvén wave amplitudes, we have performed a parametric study of the solar wind properties using our 3D MHD code. Our model provides assessment of the physical effects caused by the thermal pressure, Alfvén wave pressure and rotational forces on the wind properties for each of the wind base parameter set. To study the sensitivity of the mass loss rates and the wind velocity to changes in the coronal base temperature density, magnetic field and wave amplitude, we have calculated 12 models (the mass-loss rates and fast/slow solar wind velocities) including 9 M1 and 3 M2 models for the following range of base parameters: $T_b$ = 1.6 MK, 1.8 MK, 2 MK; $N_b$ = (25, 50, 100) x $10^7$ cm$^{-3}$; $B_0$ = 41.2, 54.9, 68.6 G ; $δV_b$ = 35, 100, 150 km/s. We also studied the effect of rotation rate on mass loss rates for stellar rotation periods between 2.5 d – 25 d.

Table 1 shows the strong effect of the wind base temperature on the calculated mass loss rates for the M1 and M2 scenarios. The temperature increase by 20% leads to a factor of 2 in mass loss rate enhancement for the M1 (young Sun) scenario and a factor of 3 increase in the mass loss rate for the M2 (the intermediate Sun's model) scenario, using the wind parameters described in Table 1. The increase of the base temperature specifies the thermal pressure gradient that presumably plays a crucial role in the wind acceleration and mass loss rates.

Our parametric study also suggests that the mass loss rate in the M1 and M2 scenarios is scaled linearly with the wind base density, $\dot{M} \propto N_b$, which is consistent with the paleo solar wind models of Sterenborg et al. (2011). The maximum mass loss rate in the M1 (at $N_b$ = $10^9$ cm$^{-3}$) scenario reaches 1.7 x $10^{-12}$ $M_\odot$/yr, where $M_\odot$ is the solar mass. We note that Sterenborg's paleo wind model required a 10 times denser wind base for comparable mass loss rates.

Our calculations find that the mass loss rate scales with the wave amplitude $δV_b$ as $δV_b^{1.15}$, which can be understood as the factor in setting the wave energy flux launched from the base. It is also important to note that the fast and slow wind velocities increase with the wave amplitude as ~ $δV_b^{0.6}$. The greater velocities for the fast and slow components are mostly driven by the increased contribution from Alfvén wave ponderomotive force (caused by the gradient of the wave pressure) with respect to the thermal pressure gradient force.

The mass loss rate does not show sensitivity to the background magnetic field set at the base, $B_0$, consistent with results reported by Sterenborg et al. (2011). For example, by increasing the field strength by 25%, the mass loss drops by about 10%. This may suggests that the slow wind emanating from the low latitude regions of the star performs work in stretching and opening magnetic field lines for the plasma to escape.

Our study does not reveal any significant effect of the solar rotation rate on the mass loss rate for the M1 scenario. This is because the thermal pressure and ponderomotive force are much greater than the gravity force and the centrifugal force, even for stars rotating with the rotation period as short as 2.5 days.

Figure 1 presents the latitudinal distribution of the wind radial velocity for the M1 (red), M2 (blue) and M3 (green) scenarios. The bi-modal wind distribution in the three evolutionary scenarios represents the result of the self-consistent solution from out model that assumes a simple dipole magnetic field. The polar open magnetic field regions are the sources of the fast wind, while the slow wind is associated with low-latitude closed field regions. For the young Sun's wind model, the fast wind velocity reaches 1372 km/s at 1 AU, while the slow wind is accelerated to 703 km/s. The M3 scenario provides the latitudinal distribution of the bi-modal solar wind that is consistent with *Ulysses* data at 1 AU (Usmanov et al. 2014). The fast wind is mostly driven by the magnetic pressure gradient provided by Alfvén waves generated at the wind base and propagating upward along a radially diverging coronal hole. For example, the M1 model shows that as waves propagate upward, they become non-linear, reaching the amplitude, $\delta B$, of 25% of the background longitudinal field at heights greater than 5 $R_\odot$. They exert ponderomotive forces (gradient of the wave pressure) on the plasma that provide the momentum for wind acceleration.

Figure 2 shows the radial profiles of the high latitude fast wind's radial velocity between the wind base and 20 $R_\odot$ for the M1 (red), M2 (blue) and M3 (green) scenarios. The figure shows that the resulting steady-state solution for the terminal wind velocity for the young Sun is a factor of 2 greater than for the current Sun.

Figure 3 shows the proton temperature in the fast wind for the M1, M2 and M3 scenarios from 0.3 to 1 AU. The figure shows that the proton temperature of the paleo solar wind is $5.5 \times 10^5$ K, which is twice as hot as that of the current solar wind.

Figure 4 presents the total mass loss rates from the fast and slow wind components calculated for the M1, M2 and M3 scenarios superimposed on the range of empirically derived mass loss rates for solar-like stars at various phases of evolution (Wood et al. 2005). This figure shows that our theoretical models at the three evolutionary phases of the Sun are consistent with the empirical mass loss rates within the uncertainties of the model input parameters discussed above.

## 4. Conclusions.

We developed three evolutionary scenarios of the solar wind using a three-fluid, fully three-dimensional global MHD model that describes the plasma dynamics from the wind base to 3 AU. These scenarios aim to represent our Sun at ages of 0.7 Gyr, 2 Gyr and its current epoch at 4.65 Gyr. Our global wind model accounts for the kinematics and thermodynamics of a single-fluid inner and a three-fluid outer solar wind. The model for the current Sun (M3) is validated by comparison with the wind parameters observed by the *WIND*, *Ulysses*, and *Voyager 2* spacecraft (Usmanov et al. 2014). This solar wind model provides the three-dimensional global description of the mass flow and turbulence parameters throughout the heliosphere for given boundary conditions at the solar coronal base.

We used three sets of input model parameters, including the base plasma density, Alfvén wave amplitude and the strength of the magnetic dipole field for each of the three solar wind evolutionary scenarios describing the 0.7, 2 and 4.65 Gyr old Sun. We also studied the effect of other "environmental" parameters, including the base temperature and rotation rate using observational constraints for stars at appropriate evolutionary states (or ages). The wind model for the young, rapidly rotating Sun, which has a magnetic dipole that is about three times stronger than that of the current epoch, produces a solar wind with the fast wind radial velocity of 1372 km/s and the slow wind velocity of 703 km/s. Its proton temperature reaches 5.5 x $10^5$ K and its total mass loss rate is as high as $85\dot{M}_\odot$ (for $N_b=10^9$ cm$^{-3}$). This suggests that the dynamic pressure from the paleo solar wind, $\dot{M} V_\infty$, is expected to be up to 170 times greater than the wind pressure measured at the Earth's magnetopause. The model of the intermediate age (~2 Gyr) Sun produces a total mass loss rate that is smaller by a factor of ~6 than that predicted from the young Sun. The fast/slow wind velocity predicted for this intermediate case is 920 km / 500 km.

To scale the wind properties with wind base parameters, we conducted a parametric study for solar wind evolutionary scenarios. Specifically, we concluded that the wind base temperature is one of the parameters to which the mass loss rate is most sensitive. The wind velocity, on the other hand, is not sensitive to the base temperature. Also, the mass loss rate scales approximately linearly with the base density and wave amplitude, while the resulting loss rates are not sensitive to the uncertainties in the magnetic field strength at the base. We find that, within the range of the stellar rotation periods between 2.5-25 days, the centrifugal and Coriolis forces are not strong enough to modify the mass loss rate and wind velocity.

Thus, our models suggest that the young solar wind was faster, more massive and hotter in its early history. The theoretical calculations of the mass loss rate appear to be in agreement with the empirically derived values for stars of various ages and with other treatments, such as the semi-empirical wind model predictions for the young Sun suggested by Cranmer and Saar (2011); Suzuki et al. (2013) and Johnstone et al. (2015). Our results will be important for realistic estimates of the wind dynamic pressure from the young Sun and other active stars on exoplanetary magnetospheres, and for assessments of the energy flux at the magnetopause, which is a crucial parameter in realistic specification of the Joule heating of their ionospheres and corresponding effects of atmospheric loss rate. The paleo wind properties derived here can be used as input for modeling the history of atmospheric loss from Mars, with insights derived from recent *MAVEN* data (Jakosky et al. 2015).

The young Sun's wind background properties are also critical in modeling the propagation of CMEs from the young Sun and their interaction with the magnetosphere of young Earth. Also, the knowledge of wind parameters is required to calculate the properties of shocks driven by paleo CMEs propagating on the background of the solar wind and the efficiency and spectrum of paleo SEP events using the Particle Acceleration and Transport in the Heliosphere (PATH) code (our work in progress; Zank et al. 2007).

This study serves as a starting point for extending our solar wind model to include the heliospheric interface region, with the goal of developing a completely physics-based cross-scale dynamical paleo heliospheric model. The extension will be based on the four-fluid global heliospheric model developed recently by Usmanov et al. (2015). The further application of our global code will provide insights into the evolution of paleo solar heliosphere and astrospheres of other solar-like stars. Such astrospheric model will also provide important validation capabilities for the empirical method proposed by Wood et al. (2005).

We thank Drs. William Danchi and Gordon Holman for a careful reading of the manuscript and suggestions for improvement. The work of A.V.U. was supported by NASA grants NNX13AR42G and NNX14AI63G to the University of Delaware. Supercomputer time allocations were provided by the NASA High-End Computing (HEC) Program awards SMD-14-4848 and SMD-15-5715 through the NASA Advanced Supercomputing (NAS) Division at the Ames Research Center and the NASA Center for Climate Simulation (NCCS) at the Goddard Space Flight Center. The authors wish to thank two anonymous referees for their comments and constructive suggestions that stimulated the authors to perform more detailed parametric study of the properties of the solar wind.

# References

Airapetian, V, Glocer, A., Danchi, W. 2015a, 18th Cambridge Workshop on Cool Stars, Stellar Systems, and the Sun, Proceedings of the conference held at Lowell Observatory, 8-14 June, 2014. Edited by G. van Belle and H.C. Harris, pp.257-268

Airapetian, V., Leake, J. E. & Carpenter, K. G. 2015b, 18th Cambridge Workshop on Cool Stars, Stellar Systems, and the Sun, Proceedings of the conference held at Lowell Observatory, 8-14 June, 2014. Edited by G. van Belle and H.C. Harris, 269.

Airapetian, V. & Cuntz, 2015, Giants of Eclipse: The ζ Aurigae Stars and Other Binary Systems, Astrophysics and Space Science Library, Volume 408. ISBN 978-3-319-09197-6. Springer International Publishing Switzerland, 123

Ayres, T. 2015, AJ, 150, 7.

Brandt, J. C., Heap, S. R. and 16 co-authors, 2001, ApJ, 121, 2173.

Cohen, O. 2011, NMRAS, 417, 2592.

Cohen, O. & Drake, J. J. 2014, 783, 55

Cranmer, S. R. & Saar, S. H. 2011 ApJ, 741, 23

Doyle, J. G. 1980, A&A, 87, 183.

Drake, J. J.; Cohen, O., Yashiro, S., Gopalswamy, N. 2013, ApJ, 764, 170.

Grießmeier, J.-M.; Stadelmann, A.; Penz, T.; Lammer, H.; Selsis, F.; Ribas, I.; Guinan, E. F.; Motschmann, U.; Biernat, H. K.; Weiss, W. W. 2004, A&A, 425, 753.

Güdel, M. 2004, The Astronomy and Astrophysics Review, Volume 12, Issue 2-3, pp. 71-237

Hassler, D. M., Rottman, G. J., Shoub, E. C., Holzer, T. E. 1990, ApJ, 348, L77.

Hollweg, J. V. 1974, JGR, 79, 3845

Hollweg, J. V. 1976, JGR, 81, 1649

Holzwarth, V. & Jardine, M. 2007, A&A 463, 11.

Jakosky, B. M. et al. 2015, Science, 350, Issue 6261, aad0210-1.

Johnstone, C. P., Güdel, M., Lüftinger, Toth, G. & Brott, I 2015, A&A 577, A27.

Kislyakova, K. G., Holmström, M., Lammer, H., Erkaev, N. V. 2015, Astrophysics and Space Science Library, Volume 411. ISBN 978-3-319-09748-0. Springer International Publishing Switzerland, 137

Khodachenko, M. L.; Alexeev, I.; Belenkaya, E.; Lammer, H.; Grießmeier, J.-M.; Leitzinger, M.; Odert, P.; Zaqarashvili, T.; Rucker, H. O. 2012, ApJ, 744, Issue 1, 70.

Kulikov, Y. N., Lammer, H., Lichtenegger, Herbert I. M.; Penz, T. Breuer, D.,Spohn, T., Lundin, R., Bienat, H. K., 2007, 129, Issue 1-3, 207

Lammer, H., Lichtenegger, H. I. M.; Kolb, C., Ribas, I., Guinan, E. F., Abart, R., Bauer, S. J. 2003, Icarus, 165, 9.

Laming, J. M., Drake, J. J., & Widding, K. G. 1995, ApJ, 443, 416.

Lundin, R., Lammer, H., Ribas, I. 2007, Planetary Magnetic Fields and Solar Forcing: Implications for Atmospheric Evolution, Space Science Reviews, Volume 129, Issue 1-3, pp. 245-278,

Matsumoto, T. & Suzuki, T. 2014, NMRAS, 440, 971.

Matt, S. P., MacGregor, K. B., Pinsonneault, M. H., & Stone, R. G. 2012, ApJ, 754, L26.

McComas, D. J. 2003, SOLAR WIND TEN: Proceedings of the Tenth International Solar Wind Conference. AIP Conference Proceedings, Volume 679, pp. 33.

McIntosh, S. W. et al. 2011, Nature, 475, 477.

Ness, J., Schmitt, J. H., M. M., Burwitz, V. Mewe, R., Raassen, A. J. J., van der Meer, R. L. J., Predehl, P., Brinkman, A. C. 2002, A&A, 394, 911.

Ness, J.-U., Güdel, M., Schmitt, J. H. M. M., Audard, M. & Telleschi, A. 2004, A&A, 427, 667.

Ofman, L. & Davila, J. M. 1998, JGR, 108 (A10), 23677.

Parker, E. N. 1958, ApJ, 128, 664.

Parker, E. N. 1979, Cosmical magnetic fields: Their origin and their activity, Oxford, Clarendon Press; New York, Oxford University Press

Pavlov, A. A., Pavlov, A. K., Mills, M. J., Ostryakov, V. M., Vasilyev, G. I. & Toon, O. B. 2005, Geophys. Res. Lett. , 32, Issue 1, L01815.

Pizzo, V. J. 1978, JGR, 83, 5563.

Pizzo, V. J. 1982, JGR, 87, 4374.

Potgieter, M. S. 2013, Living Rev. Solar Phys., 10, 3.

Ribas, I., Guinan, E. F., Gudel, M., & Audard, M. 2005, ApJ, 622, 680.

Schrijver, C. J., Dobbins, R., Murtagh, W., Petrinec, S. M. 2014, Space Weather, 12, 487.

Skumanich, A. 1972, ApJ, 171, 565.

Sterenborg, M. G., Cohen, O. , Drake, J. J., Gombosi, T. I. 2011, 116, A01217.

Suzuki, T. K., Imada, S., Kataoka, R., Kato, Y., Matsumoto, T., Miyahara, H., Tsuneta, S. 2013, PASJ, 65, 98

Usmanov, A. 1993, Sol. Phys. 146, 377.

Usmanov, A. V., Goldstein, M. L., Besser, B. P., Fritzer, J. M. 2000, JGR, 105, A6, 12675.

Usmanov, A. V., Goldstein, M. L., Matthaeus, W. H. 2014, ApJ, 788, 43.

Usmanov, A. V., Goldstein, M. L., Matthaeus, W. H. 2015, ApJ, submitted.

Usoskin I. G. & Kovaltsov, G. A. 2008, C. R. Geoscience, 340, 441.

van der Holst, B., Manchester IV, W.B., Frazin, R.A., Vaśquez, A.M., To th, G. and Gombosi, T.I., 2010, ApJ, 725, 1373.

Vidotto, A. A., Opher, M., Jatenco-Pereira, V., Gombosi, T. I. 2009, ApJ, 699, 441.

Vidotto, A. A., Gregory, S. G., Jardine, M., Donati, J. F., Petit, P., Morin, J., Folsom, C. P., Bouvier, J., Cameron, A. C., Hussain, G. and 5 co-authors 2014, NMRAS, 441, 2361

Y. Wei, M. Fraenz, E. Dubinin, J. Woch, H. Lühr, W. Wan, Q.-G. Zong, T. L. Zhang, Z. Y. Pu, S. Y. Fu, S. Barabash, R. Lundin, and I. Dandouras 2012, JGR, 117, A03208.

Wood, B. E., Linsky, J. L., Ayres, T. R. 1997, ApJ, 478, 745.

Wood, B. E., Müller, H-R, Zank, G. P., Linsky, J. L. & Redfield, S. 2005, ApJ, 628, L143.

Zank, G. P., Li, G. & Verkhoglyadova, O. 2007, Space Sci. Rev., 130, 11.

Zhao, X. & Hoeksema, J. T. J. Geophys. Res., 1995, 100, 19

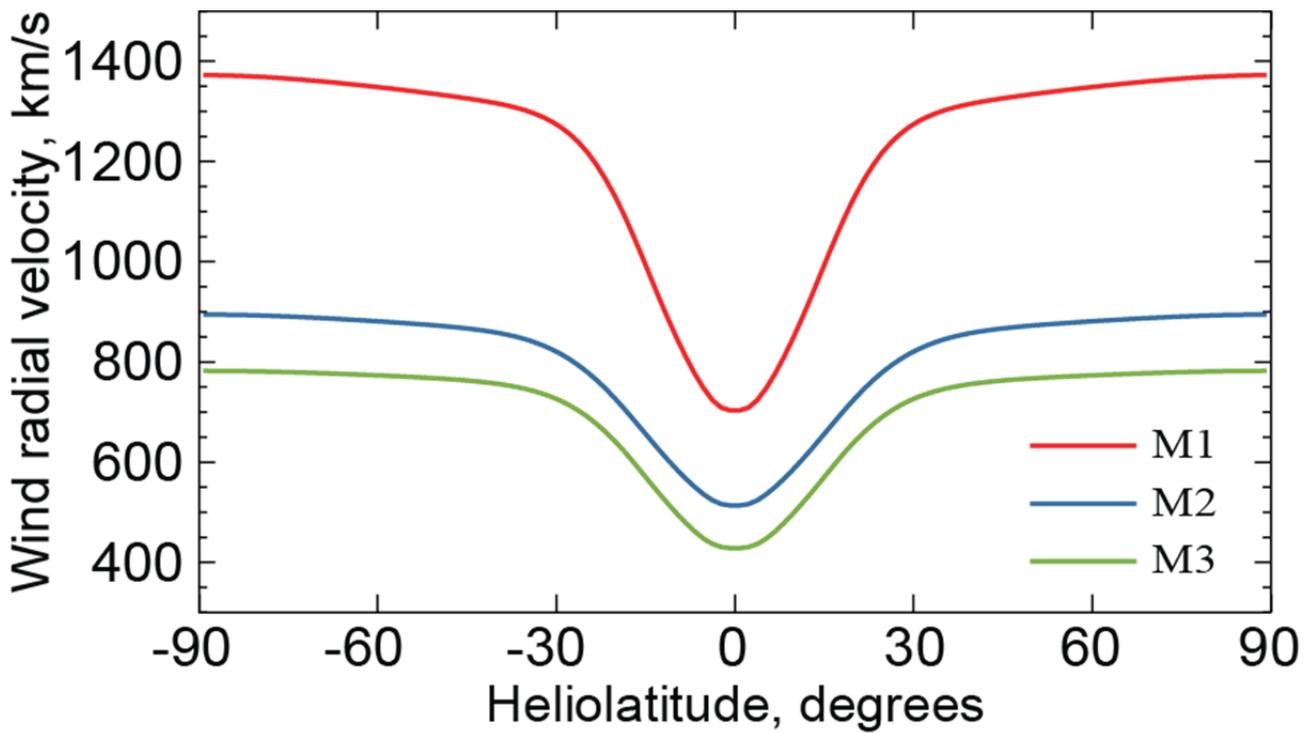

**Figure 1**. The latitudinal distribution of the solar wind radial velocity at 1 AU. The red, blue and green curves represent the wind's velocity profiles from the young Sun at 0.7 Gyr (M1 scenario), the intermediate age Sun at 2 Gyr (M2 scenario) and the current Sun (M3 scenario) respectively.

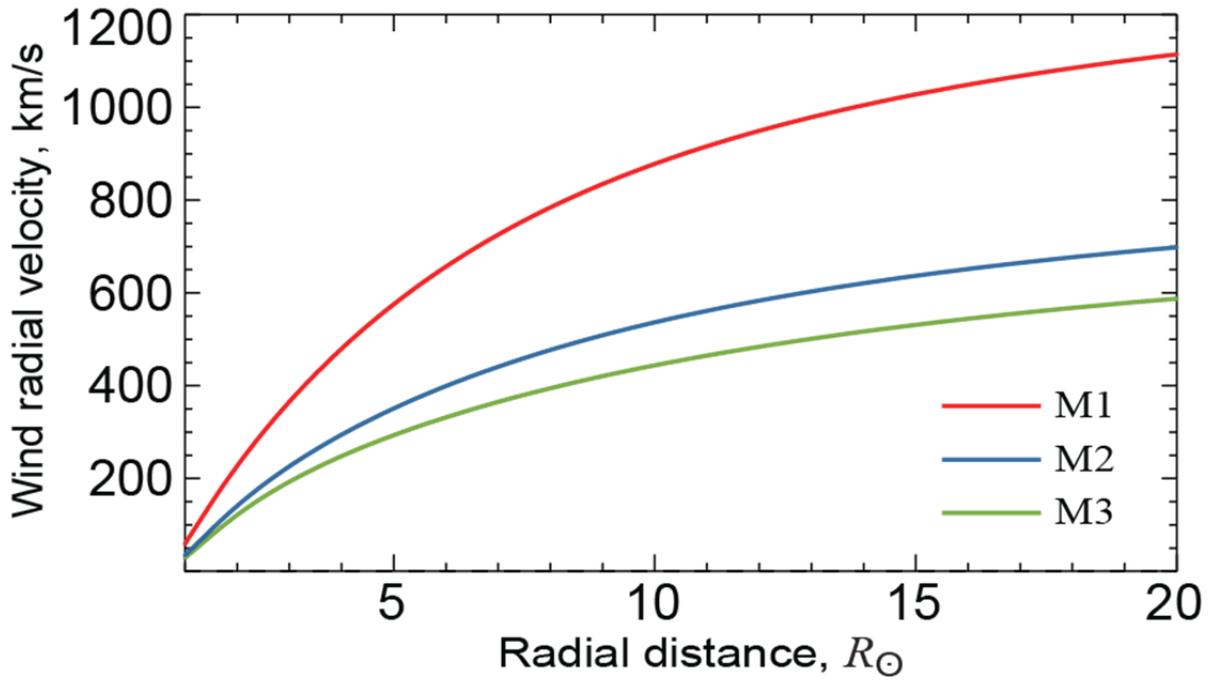

**Figure 2**. Radial profiles of the high altitude fast solar wind radial velocity. The red, blue and yellow curves describe the wind's velocity profile from the wind base to 20 $R_\odot$ for the models of the young Sun at 0.7 Gyr (M1 scenario), the intermediate age Sun at 2 Gyr (M2 scenario) and the current Sun (M3 scenario), respectively.

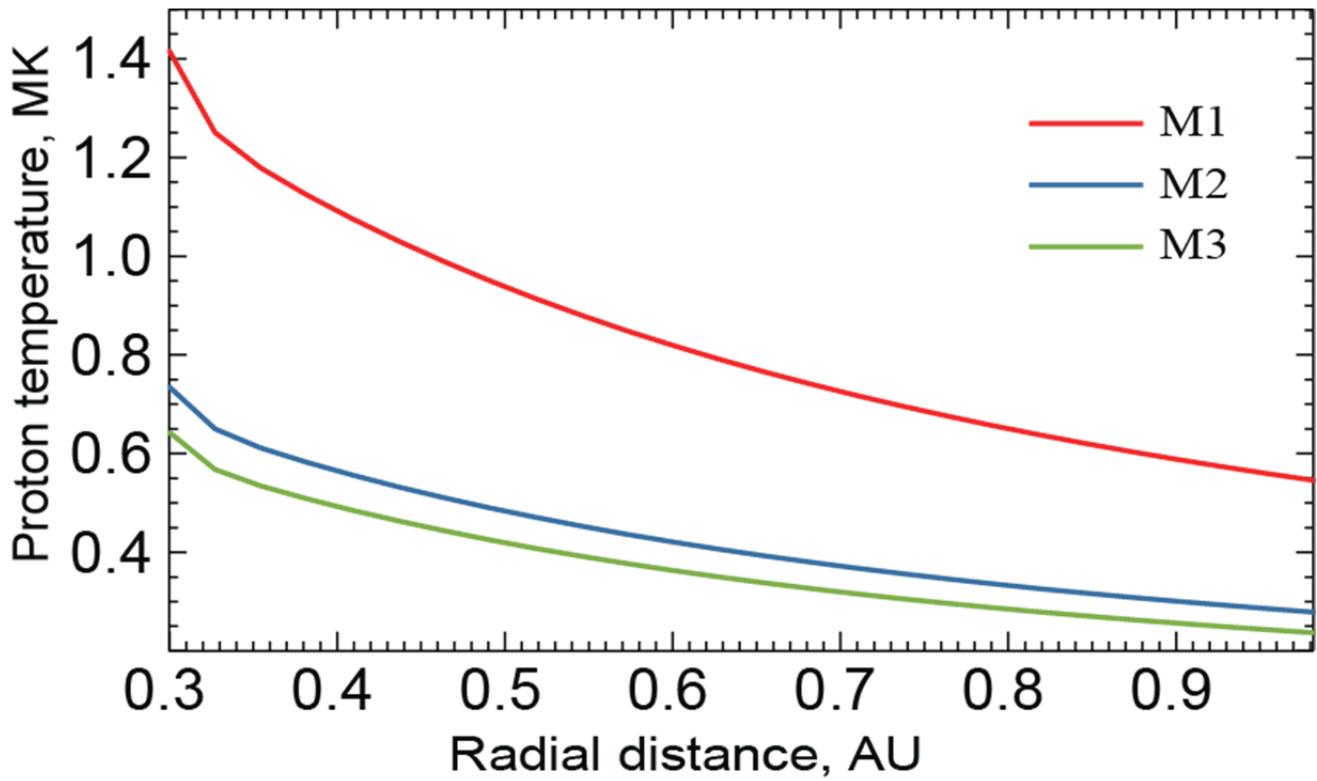

**Figure 3**. Radial profiles of the proton temperature in the fast wind. The red, blue and green curves describe the wind's temperature profiles for the models of the young Sun at 0.7 Gyr (M1 scenario), the intermediate age Sun at 2 Gyr (M2 scenario) and the current Sun (M3 scenario), respectively.

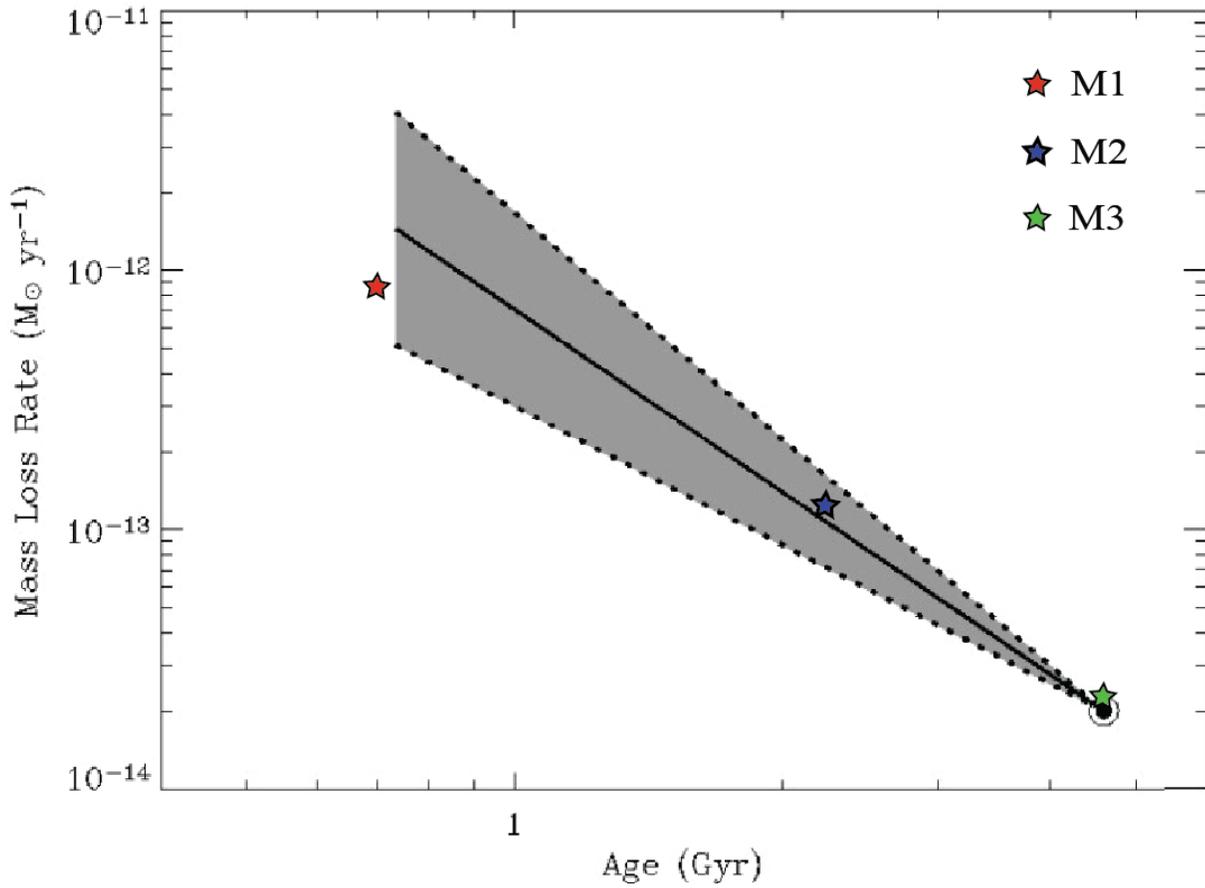

**Figure 4**. The total mass-loss rates from Alfvén-wave driven solar winds at 0.7 Gyr (red star), 2 Gyr (blue star) and 4.6 Gyr (green star) superimposed on the empirically derived values of mass loss rates (grey area) from a sample of solar-type stars of various ages (Wood et al. 2005).